\documentclass[iop,apjl]{emulateapj}
\usepackage[varg]{txfonts}
\newcommand{\se}[1]{Section\ \ref{sec:#1}}

\newcommand{\eq}[1]{Equation\ (\ref{eq:#1})}

\newcommand{\eqs}[2]{Equations\ (\ref{eq:#1}) and (\ref{eq:#2})}

\newcommand{\eqsto}[2]{Equations\ (\ref{eq:#1})--(\ref{eq:#2})}

\newcommand{\Eqs}[2]{Equations\ (\ref{eq:#1}) and (\ref{eq:#2})}

\newcommand{\fg}[1]{Figure\ \ref{fig:#1}}

\newcommand{\Fg}[1]{Figure\ \ref{fig:#1}}
\newcommand{\Tb}[1]{\mbox{Table\ \ref{tab:#1}}}

\newcommand{\eg}{e.g.,}

\newcommand{\cf}{cf.}
\newcommand{\micr}{\ensuremath{\mu\mathrm{m}}}
\newcommand{\mEarth}{\ensuremath{M_\oplus}}

\newcommand{\mast}{\ensuremath{m^\ast}}
\newcommand{\Mdotcum}[1]{\ensuremath{\dot{M}_\mathrm{#1}}}
\newcommand{\Mdottot}[1]{\ensuremath{\dot{M}_\mathrm{#1}}}

\newcommand{\dMdrF}[1]{\ensuremath{\frac{d\dot{M}_\mathrm{#1}}{dr}}}
\newcommand{\dMdrf}[1]{\ensuremath{d\dot{M}_\mathrm{#1}/dr}}

\usepackage{color}
\usepackage{framed}

\shorttitle{An atmospheric structure equation for grain growth}
\shortauthors{Ormel}

\begin{document}

%% LaTeX will automatically break titles if they run longer than
%% one line. However, you may use \\ to force a line break if
%% you desire.

\title{An atmospheric structure equation for grain growth}

%% Use \author, \affil, and the \and command to format
%% author and affiliation information.
%% Note that \email has replaced the old \authoremail command
%% from AASTeX v4.0. You can use \email to mark an email address
%% anywhere in the paper, not just in the front matter.
%% As in the title, use \\ to force line breaks.

\author{C.W. Ormel\altaffilmark{1}}
\affil{Astronomy Department, University of California, Berkeley, CA 94720}
\email{ormel@astro.berkeley.edu}

%\author{S. Okuzumi\altaffilmark{2,3}}
%\affil{Department of Earth and Planetary Sciences, Tokyo Institute of Technology, Meguro-ku, Tokyo, 152-8551} 
%\email{okuzumi@geo.titech.ac.jp}

%% Notice that each of these authors has alternate affiliations, which
%% are identified by the \altaffilmark after each name.  Specify alternate
%% affiliation information with \altaffiltext, with one command per each
%% affiliation.

\altaffiltext{1}{Hubble Fellow}
%\altaffiltext{2}{JSPS Superlative Research Fellow}
%\altaffiltext{3}{Department of Physics, Nagoya University, Nagoya, Aichi 464-8602, Japan}
%% Mark off your abstract in the ``abstract'' environment. In the manuscript
%% style, abstract will output a Received/Accepted line after the
%% title and affiliation information. No date will appear since the author
%% does not have this information. The dates will be filled in by the
%% editorial office after submission.

%Scattering of planetesimals causes a planet to migrate.  
\begin{abstract}
We present a method to include the evolution of the grain size and grain opacity $\kappa_\mathrm{gr}$ in the equations describing the structure of protoplanetary atmospheres. The key assumption of this method is that a single grain size dominates the grain size distribution at any height $r$. In addition to following grain growth, the method accounts for mass deposition by planetesimals and grain porosity. We illustrate this method by computation of a simplified atmosphere structure model. In agreement with previous works, grain coagulation is seen to be very efficient. The opacity drops to values much below the often-used `ISM-opacities' ($\sim$$1\ \mathrm{cm^2\ g}^{-1}$) and the atmosphere structure profiles for temperature and density resemble that of the grain-free case. Deposition of planetesimals in the radiative part of the atmosphere hardly influences this outcome as the added surface is quickly coagulated away. We observe a modest dependence on the internal structure (porosity), but show that filling factors cannot become too large because of compression by gas drag.
\end{abstract}

%% Keywords should appear after the \end{abstract} command. The uncommented
%% example has been keyed in ApJ style. See the instructions to authors
%% for the journal to which you are submitting your paper to determine
%% what keyword punctuation is appropriate.


%% From the front matter, we move on to the body of the paper.
%% In the first two sections, notice the use of the natbib \citep
%% and \citet commands to identify citations.  The citations are
%% tied to the reference list via symbolic KEYs. The KEY corresponds
%% to the KEY in the \bibitem in the reference list below. We have
%% chosen the first three characters of the first author's name plus
%% the last two numeral of the year of publication as our KEY for
%% each reference.

%% Authors who wish to have the most important objects in their paper
%% linked in the electronic edition to a data center may do so by tagging
%% their objects with \objectname{} or \object{}.  Each macro takes the
%% object name as its required argument. The optional, square-bracket 
%% argument should be used in cases where the data center identification
%% differs from what is to be printed in the paper.  The text appearing 
%% in curly braces is what will appear in print in the published paper. 
%% If the object name is recognized by the data centers, it will be linked
%% in the electronic edition to the object data available at the data centers  
%%
%% Note that for sources with brackets in their names, e.g. [WEG2004] 14h-090,
%% the brackets must be escaped with backslashes when used in the first
%% square-bracket argument, for instance, \object[\[WEG2004\] 14h-090]{90}).
%%  Otherwise, LaTeX will issue an error. 

\section{Introduction}
\label{sec:intro}
Once (proto)planets reach sizes of $\sim$$10^3$ km they start to bind the gas of the disk, forming an atmosphere. The evolution of these atmospheres is usually modeled by solving the 1D stellar structure equations \citep[\eg][]{PollackEtal1996,PapaloizouTerquem1999,IkomaEtal2000,Rafikov2006,AlibertEtal2005,HubickyjEtal2005,FortierEtal2007,MordasiniEtal2009,MordasiniEtal2014,PisoYoudin2014}. A major source of uncertainty in these works concerns the adopted value of the grain opacity $\kappa_\mathrm{gr}$. Traditionally, following \citet{Stevenson1982}, large, ISM-like values ($\sim$$1\ \mathrm{cm^2\ g}^{-1}$), are adopted. A much lower $\kappa_\mathrm{gr}$, however, allows heat to escape more efficiently, causing the atmosphere to contract and the densities to rise. In the grain-free limit the atmospheres mass may already collapse at core masses of $\sim$1 $\mEarth$ \citep{HoriIkoma2010}. Clearly, $\kappa_\mathrm{gr}$ matters and there is a desire to follow the evolution of the grain size in these atmospheres.

Grain growth (coagulation) and settling will reduce the opacity. These effects are sometimes accounted for by an arbitrary reduction of $\kappa_\mathrm{gr}$ with respect to the ISM-values, which is clearly ad-hoc. A much preferred approach in terms of accuracy is to solve the \citet{Smoluchowski1916} coagulation equation \citep{Podolak2003,MovshovitzPodolak2008,MovshovitzEtal2010,RogersEtal2011}. Hower, this has the drawback of increasing the complexity of the model -- and the computational expense -- as it adds an extra dimension. 
%The formal way to proceed is to solve the  for the  . However, this has the drawback of . 
Here, we will present an approximate method that solves for the \textit{characteristic grain size} $s$ as function of atmosphere depth $r$.  Our method entails solving an ordinary differential equation (ODE), in addition to the ODEs for pressure, temperature, and luminosity. 

The advantage of our approach is that it is far more realistic than simply assuming a constant $\kappa_\mathrm{gr}$ but that it avoids the computationally intensive calculations of solving for the grain size distribution. The high `bang for the buck' of our approach facilitates running a vast parameter study. The method can be readily incorporated in the machinery of the above works as well as be applied to planet population synthesis codes.

We present this method in \se{method} and apply it to an atmosphere model in \se{results}.  The atmosphere model is intentionally simple as the goal of this paper is to illustrate the implications of (neglecting) grain coagulation. In \se{summary} we summarize our findings.

\section{The method}
\label{sec:method}
\subsection{The idea}
The fundamental assumption of this method is that the grain size distribution at any height $r$ is characterized by a single mass $m^\ast$ or equivalently its corresponding radius $s$. The size $s$ should corresponds to the particles that dominate the mass budget of the distribution.

%We identify $m^\ast$ as the particle that carries most of the mass of the distribution. 
Such characteristic size method have been used successfully to follow the grain growth in the protoplanetary disk (\citealt{BirnstielEtal2012i}; S.\ Okuzumi 2014, pers.\ comm). A possible caveat is that deposition of small grains by planetesimal breakup renders the grain size distribution bimodal, as seen in Fig.\ 4 of \citet{MovshovitzPodolak2008}. We describe a correction for this bimodality in \se{bimodal}.
%, but find that it has little effect. 
%observed bimodality in the deeper layers of the atmosphere (see their Fig.\ 4).  However, in their case it appears that the large grains still dominate the opacity (also by virtue of the fact that small grains are inefficient absorbers). 
Recently, \citet{Mordasini2014}, also applying the characteristic grain assumption, derived an analytical expression for $\kappa_\mathrm{gr}$. His model is cruder than ours yet compares favorably with the detailed calculations of \citet{MovshovitzPodolak2008}, supporting the viability of the characteristic grain approximation. 
%It is certainly preferable over fixing $\kappa_\mathrm{gr}$.

A further refinement of the method (not implemented here) is to solve for the power-law of the size distribution for masses $m<\mast$ \citep[\cf][]{EstradaCuzzi2008,BirnstielEtal2011} to additionally obtain the size where the opacity peaks (if different from $s$).

\subsection{Formulation}
The transport equation for the grain density reads:
\begin{equation}
  \label{eq:transport}
  \frac{\partial \rho_\mathrm{gr}}{\partial t}
  = \nabla \cdot (D \nabla \rho_\mathrm{gr}) 
    -\nabla \cdot (\mathbf{v}X) 
    +\dot\rho_\mathrm{dep},
\end{equation}
where the terms on the right hand side account for diffusion, settling, and deposition of grains.
%and the second for the settling, and the source (and sink) term accounts for processes as grain growth, grain deposition by disintegrating planetesimals, grain evaporation, and grain nucleation.  
For simplicity, we ignore grain diffusion in this work, $D=0$. Diffusion is more important for the convective regions of the atmosphere, where however the (grain) opacity no longer matters. Without diffusion, transport of grains is always downwards at a settling velocity, $\mathbf{v} = -v_\mathrm{settl}(\mast)\mathbf{e}_r$, where $v_\mathrm{settl}$ is a function of the grain aerodynamical properties and the local gravitational acceleration $g_r$. 

In our model we consider only mass deposition by disintegrating planetesimals as a source for the grain density.\footnote{It is straightforward to extend the model with processes as grain vaporization and nucleation.} Let $\Mdotcum{dep}(r)$ be the \textit{cumulative} mass flux of solids that have disintegrated into small grains by radius $r$. At the top of the atmosphere $\Mdotcum{dep}$ equals the accretion rate due to small grains captured from the disk, $\dot{M}_\mathrm{disk}$. It increases inwards due to deposition of grains from disintegrating planetesimals to equal the total total accretion rate $\Mdotcum{tot}$ at the core radius $r_\mathrm{core}$. %The difference is due to desintegrating planetesimals in the atmosphere, $\Mdotcum{P}(r)=\Mdotcum{dep}(r)-\Mdottot{disk}$.
%Let $\Mdotcum{P}(r)=\Mdotcum{dep}-\Mdottot{disk}$ represent the planetesimal contribution. 
See \fg{deposit} where we envisioned that planetesimals breakup around a radius $r_\mathrm{crit}$. The mass in grains that planetesimals deposit in a shell $[r,r+\Delta r]$ is thus $-(\dMdrf{dep})\Delta r$. The spatial density of grains then increases as:
\begin{equation}
  \dot\rho_\mathrm{dep}
  =
  -\frac{1}{4\pi r^2} \dMdrF{dep};
  \label{eq:drhodt-dep}
\end{equation}
and the transport equation becomes
\begin{equation}
  \frac{\partial\rho_\mathrm{gr}}{\partial t}
  = 
  \frac{1}{r^2} \frac{\partial}{\partial r} \left( r^2 v_\mathrm{settl} \rho_\mathrm{gr} \right)
  -\frac{1}{4\pi r^2} \frac{d\dot{M}_\mathrm{dep}}{dr}.
  \label{eq:master-1}
\end{equation}

\begin{figure}[t]
  \includegraphics[width=88mm]{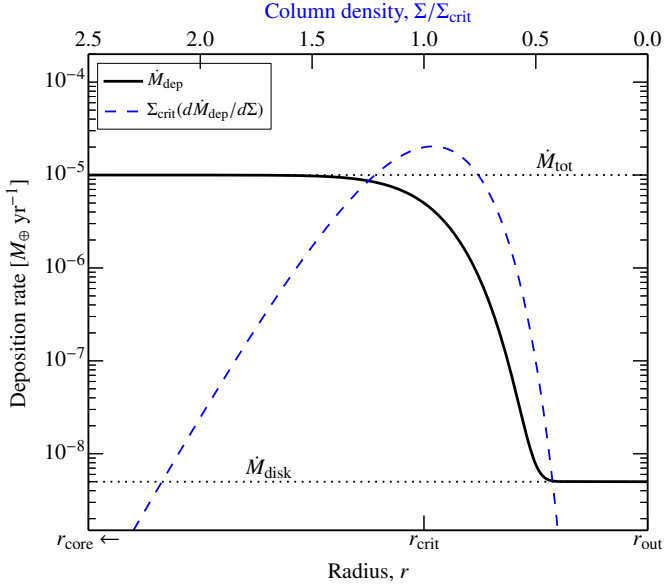}
  \caption{The deposition profile ($\Mdotcum{dep}$, solid) used in the calculations of \se{results}. It consists of a constant contribution from small grains $\Mdotcum{disk}$ and of an $r$-dependent contribution from planetesimal breakup around a radius $r_\mathrm{crit}$. The differential deposition profile, plotted here as function of column density (top axis), is described in \se{depos}. Note that $-\dMdrf{dep}=\rho_\mathrm{gas}(d\dot{M}_\mathrm{dep}/d\Sigma)$.}
  \label{fig:deposit}
% Planetesimals fragment and deposit small grains at a characteristic surface density $\Sigma_\mathrm{crit}$. The solid line shows the deposition profile, which is here modeled by a log-normal distribution with shape parameter $\sigma=0.1$. }
\end{figure}
Grain growth does not affect the average density $\rho_\mathrm{gr}$. However, coagulation increases the characteristic mass $\mast$ on a timescale $T_\mathrm{grow}$:
\begin{equation}
  \left. \frac{\partial\mast}{\partial t} \right)_\mathrm{grow}
  = \frac{\mast}{T_\mathrm{grow}}.
  \label{eq:Tgrow}
\end{equation}
Grain deposition also affects $\mast$, driving it towards the mass of the deposited grains $m_\mathrm{dep}$. The rate at which this occurs depends on the deposition rate and on the density in $m^\ast$ grains. If only grain deposition affects $\mast$:
\begin{equation}
  m^\ast (t+\Delta t)
  = \frac{\rho_\mathrm{gr}\mast +m_\mathrm{dep}\dot\rho_\mathrm{dep}\Delta t}{\rho_\mathrm{gr} +\dot\rho_\mathrm{dep}\Delta t}.
  \label{eq:mast-time}
\end{equation}
The weighing with the density in \eq{mast-time} reflects the fact that $\mast$ follows the mass of the distribution.\footnote{Formally, $\mast$ can be defined as the ratio of the second to first moment of the grain density distribution function $n_\mathrm{gr}(m,t)$.} Taking $\Delta t\rightarrow 0$ we obtain the rate at which $\mast$ changes:
\begin{equation}
  \left. \frac{\partial\mast}{\partial t} \right)_\mathrm{dep}
  = \frac{\dot\rho_\mathrm{dep}}{\rho_\mathrm{gr}} (m_\mathrm{dep} -\mast),
  \label{eq:dmastdt}
\end{equation}
showing that the shift of $\mast$ towards $m_\mathrm{dep}$ speeds up when the planetesimal mass deposition is large and the grain density low.

The grain characteristic mass evolves according to
\begin{equation}
  \frac{D\mast}{Dt} 
  =
  \frac{\partial\mast}{\partial t}
  - v_\mathrm{settl} \frac{\partial \mast}{\partial r}
  = 
  \textrm{source terms}
  %\left. \frac{\partial\mast}{\partial t} \right)_\mathrm{grow}
  %+ \left. \frac{\partial\mast}{\partial t} \right)_\mathrm{dep}
\end{equation}
where $D/Dt$ is the Lagrangian derivative. With \eqs{Tgrow}{dmastdt} as source terms:
\begin{equation}
  \frac{\partial\mast}{\partial t}
  = v_\mathrm{settl} \frac{\partial \mast}{\partial r}
  +\frac{\mast}{T_\mathrm{grow}}
  -\frac{m_\mathrm{dep}-\mast}{4\pi \rho_\mathrm{gr} r^2} \frac{d\dot{M}_\mathrm{dep}}{dr}.
  \label{eq:master-2a}
\end{equation}

\subsection{Steady-state equations}
The expressions greatly simplify when a steady state can be assumed, $\partial/\partial t=0$. A requirement for a steady solution is that the grain transport timescale is short compared to the timescale on which the density and temperature structure of the atmosphere evolve, and to changes in $\Mdotcum{dep}(t)$. We will show that grain settling times are $\lesssim$$10^3$ yr, which validates the assumption.

In that case, \eq{master-1} integrates into
\begin{equation}
  \rho_\mathrm{gr} = \frac{\dot{M}_\mathrm{dep}(r)}{4\pi r^2 v_\mathrm{settl}}
  \label{eq:master-4a}
\end{equation}
which expresses mass conservation. Using this equation, the steady state version of \eq{master-2a} reads:
\begin{equation}
  \frac{\partial\mast}{\partial r}
  = -\frac{\mast}{v_\mathrm{settl}T_\mathrm{grow}}
  +\frac{m_\mathrm{dep}-\mast}{\dot{M}_\mathrm{dep}} \frac{d\dot{M}_\mathrm{dep}}{dr}.
  \label{eq:master-4b}
\end{equation}
This is an ordinary differential equation (ODE) for the characteristic mass $\mast$. It supplements the atmospheric structure equations for pressure, temperature, and luminosity.

\subsection{Bimodal extension}
\label{sec:bimodal}
In steady state we can calculate the density of $m_\mathrm{dep}$-grains:
\begin{equation}
  \rho_\mathrm{dep}
  = \dot\rho_\mathrm{dep} T_\mathrm{sweep},
\end{equation}
where $T_\mathrm{sweep}$ is the timescale for the $m_\mathrm{dep}$-grains to be swept-up by the $m^\ast$-grains: $T_\mathrm{sweep} = 1/(\pi s^2 v_\mathrm{settl}\rho_\mathrm{gr}/m^\ast)$. In such a two component model \eq{dmastdt} no longer applies, but is replaced as source term with:
\begin{equation}
  \left. \frac{\partial m^\ast}{\partial t} \right)_\mathrm{sweep}
  = \frac{\dot\rho_\mathrm{dep}m^\ast}{\rho_\mathrm{gr}}.
  \label{eq:dmdt-sweep}
\end{equation}

\section{Results}
\label{sec:results}
\subsection{Atmosphere structure equations}
\label{sec:atmos}
As an illustration of \eqs{master-4a}{master-4b}, we compute an atmosphere structure model. In contrast to the works mentioned in the Introduction our model is extremely rudimentary: the sole aim of our idealized model is to explore the effects of grain growth and grain settling. We occasionally provide references at points where the model can be extended.

The following setup is considered: a $M_\mathrm{core} = 5\ \mEarth$ core at 5.2 AU accreting solids (planetesimals and grains) at a rate of $\dot{M}_\mathrm{tot} = 10^{-5} \mEarth\ \mathrm{yr}^{-1}$. These and other model parameters are listed in \Tb{list}. 
We assume that the planet atmosphere is static and that the luminosity $L$ entirely originates from the planetesimals and their collisional products that rain down on the core. We ignore self-gravity. In that case the atmospheric structure equations read:
\begin{eqnarray}
  \label{eq:P-eqn}
  \frac{\partial P}{\partial r}
  &=& - G M_\mathrm{core}\frac{\rho_\mathrm{gas}}{r^2} \\
  \label{eq:T-eqn}
  \frac{\partial T}{\partial r}
  &=& -\frac{\partial P}{\partial r} \frac{T}{P} \nabla
\end{eqnarray}
where $P$ is pressure, $T$ temperature, $\rho_\mathrm{gas}$ gas density, and $G$ Newton's gravitational constant. The thermal gradient is $\nabla = \min(\nabla_\mathrm{rad}, \nabla_\mathrm{ad})$ with $\nabla_\mathrm{ad}$ the adiabatic gradient and $\nabla_\mathrm{rad}$ the radiative gradient:
\begin{equation}
  \nabla_\mathrm{rad}
  = -\frac{3\kappa L}{64\pi\sigma_\mathrm{sb}GM_\mathrm{core}} \frac{P}{T^4},
\end{equation}
where $\kappa$ is the opacity (in $\mathrm{cm}^2$ per unit gram gas) and $\sigma_\mathrm{sb}$ Stefan-Boltzmann constant. The luminosity $L$ generated by the impacting planetesimals is given by $L=GM_\mathrm{core} \dot{M}_\mathrm{tot}/r_\mathrm{core}$. \Eqs{P-eqn}{T-eqn} are supplemented by the ideal equation of state: 
\begin{equation}
  \label{eq:EOS}
  P = \frac{\rho_\mathrm{gas} k_B T}{\mu},
\end{equation}
where $\mu$ is the mean molecular weight and $k_B$ Boltzmann's constant.

%Under these simplifying assumptions, 
The opacity $\kappa$ in $\nabla_\mathrm{rad}$ is the sum of the gas and grain opacities:
\begin{equation}
  \kappa =
  \kappa_\mathrm{gas} +\kappa_\mathrm{gr}
  = \kappa_\mathrm{gas} +\kappa_\mathrm{geom}Q_e,
\end{equation}
where the geometrical opacity follows from the grain abundance $Z_\mathrm{gr}$ and characteristic size $s$: $\kappa_\mathrm{geom}=3Z_\mathrm{gr}/4\rho_\bullet s$, with $Z_\mathrm{gr}=\rho_\mathrm{gr}/\rho_\mathrm{gas}$, $\rho_\bullet$ the grain internal density, and $Q_e$ the efficiency factor. The gas opacity in atmosphere structure models is usually provided by lookup tables \citep{FergusonEtal2005,FreedmanEtal2008}. For grain opacities, approximate recipes have recently been published \citep{KataokaEtal2013ii,CuzziEtal2014}, which provide $\kappa_\mathrm{gr}$ for general grain properties (composition, sizes, internal structure) without the need for Mie calculations. For the purposes of this paper it suffices to use crude analytical expressions: $\kappa_\mathrm{gas}=10^{-8}\rho_\mathrm{gas}^{2/3} T^3$ (cgs-units; \citealt{BellLin1994}) and $Q_e = \min(0.3x,2)$ with $x=2\pi s/\lambda_\mathrm{max}$ and $\lambda_\mathrm{max}(T)$ the peak wavelength from Wien's displacement law.

\begin{deluxetable}{lp{44mm}r}
%\begin{deluxetable}{lp{84mm}r}
  \tablecaption{\label{tab:list}Model parameters and description}
  \tablehead{ Parameter & Description & Value}
  \startdata
    $E_\mathrm{roll}$       & Rolling energy             & $1\times10^{-8}$ ergs   \\
    $M_\mathrm{core}$       & Core mass                  & $5\ \mathrm{M_\oplus}$   \\
    $\dot{M}_\mathrm{tot}$  & Total solid accretion rate & $10^{-5}\ \mathrm{M_\oplus\ yr}^{-1}$ \\
    $\dot{M}_\mathrm{disk}$ & Disk contribution to the solid accretion rate & $5\times10^{-9}\ \mathrm{M_\oplus\ yr}^{-1}$ \\
    $T_\mathrm{disk}$       & Disk temperature          & 150 K \\
%   $g_r$                   & planet gravity \\ 
    $a_\mathrm{disk}$       & Disk orbital radius of the planet & 5.2 AU \\
    $m_\mathrm{dep}$        & Mass of the deposited grains  & monomer mass \\
    $r_\mathrm{Bondi}$      & Bondi radius              & $3.7\times10^{11}\ \mathrm{cm}$  \\
    $r_\mathrm{core}$       & Core radius               & $1.2\times10^9\ \mathrm{cm}$  \\
    $r_\mathrm{out}$        & Outer atmosphere radius (= Hill radius) & $1.3\times10^{12}\ \mathrm{cm}$ \\
    $s_0$                   & (Monomer) grain radius      & $1\ \micr$ \\
    $\nabla_\mathrm{ad}$    & Adiabatic temperature gradient & 0.28 \\
    $\Sigma_\mathrm{crit}$  & Characteristic column density where planetesimals are deposited & $10^2\ \mathrm{g\ cm}^{-2}$ \\
    $\delta$                & Fractal exponent used in porous models  & $0.8$ \\
    $\mu$                   & Mean molecular mass       & $2.34 m_\mathrm{H}$ \\
    $\rho_0$                & Monomer grain internal density & $3\ \mathrm{g\ cm^{-3}}$\\
    $\rho_\mathrm{core}$    & Planet core internal density     & $4\ \mathrm{g\ cm^{-3}}$ \\
    $\rho_\mathrm{disk}$    & Disk density              & $10^{-11}\ \mathrm{g\ cm^{-3}}$ \\
    $\sigma$                & Shape parameter determining the planetesimal mass deposition profile & 0.2 \\
    $\chi$                  & Differential drift dispersion factor  & 0.1
  \enddata
\end{deluxetable}
\begin{deluxetable*}{lp{38mm}llllll}
% \rotate
  \tabletypesize{\scriptsize}
  \tablecaption{\label{tab:runs}Model runs and results}
  \tablehead{ Name & Description & \multicolumn{3}{c}{Atmosphere mass [$\mEarth$]\tablenotemark{a}} & \multicolumn{3}{c}{Settling time [yr]\tablenotemark{b}}\\
  &             & $M_<^\mathrm{RBC}$ & $M_<^\mathrm{Bondi}$ & $M_<^\mathrm{out}$ & $T_\mathrm{settl}^\mathrm{RCB}$  & $T_\mathrm{settl}^\mathrm{Bondi}$ & $T_\mathrm{settl}^\mathrm{out}$ 
  }
  \startdata
  ISM-like                      & Fixed grain abundance $Z_\mathrm{gr}=10^{-2}$                      & $1.7\times10^{-4}$ &$1.1\times10^{-3}$    & $1.9\times10^{-2}$  & 710             & $2.5\times10^3$ & $3.6\times10^4$ \\
  Virtually grain-free          & Fixed grain abundance $Z_\mathrm{gr}=10^{-8}$                      & $3.0$            & $3.1$                  & $3.1$               & $1.3\times10^6$ & $1.3\times10^6$ & $1.4\times10^6$ \\
  Grain growth                  & Grain coagulation and settling without planetesimal breakup       & $7.3\times10^{-2}$  & $8.0\times10^{-2}$  & $9.9\times10^{-2}$  & 3.3  & 260 & $6.5\times10^3$ \\
  Grain growth and deposition   & Includes planetesimal breakup in radiative part of the atmosphere & $5.6\times10^{-2}$  & $6.1\times10^{-2}$  & $7.9\times10^{-2}$  & 0.09 & 190 & $6.4\times10^3$ \\
  Fractal growth                & Assumes grain growth is fractal                                   & $4.7\times10^{-3}$  & $6.8\times10^{-3}$  & $2.5\times10^{-2}$ & 0.04 & 340 & $1.3\times10^4$ \\
  Equilibrium $\phi$            & Assumes grain porosity is limited by gas drag                     & $0.16$           & $0.17$                 & $0.19$  & $0.11$ & $142$ & $1.2\times10^4$
  \enddata
  \tablenotetext{a}{Gas mass enclosed within the radiative-convective boundary (RBC), Bondi radius, and the outer (Hill) radius, respectively.}
  \tablenotetext{b}{Defined as $T_\mathrm{settl}^X=\int_{r_\mathrm{core}}^{r_X} dr'/v_\mathrm{settl}(r')$.}
\end{deluxetable*}
\subsection{Model summary}
\eqsto{master-4a}{T-eqn} form a system of ODEs with the radius $r$ as the independent parameter and $P$, $T$, $Z_\mathrm{gr}$ (a proxy for $\rho_\mathrm{gr}$), and $\mast$ the unknowns. We integrate from outside-in, starting at the Hill radius of the planet where the disk values for $P$ and $T$ apply. \Tb{runs} and \fg{cool} present the results. In these the bimodal extension (\se{bimodal}) is \textit{not} implemented.

\subsection{No grain growth and fixed $Z_\mathrm{gr}$}
We start with two runs that have a fixed grain radius $s=s_0=1\ \micr$ and a fixed $Z_\mathrm{gr}$ throughout the atmosphere. Figures \ref{fig:cool}a and b show the results for the `virtual grain-free' $Z_\mathrm{gr}=10^{-8}$ run (solid curves) and the `ISM-like' $Z_\mathrm{gr}=10^{-2}$ (dashed curves). The left panel (\fg{cool}a) gives the temperature and density profiles. Note that the one-third power of density is plotted.

Clearly, the value of $Z_\mathrm{gr}$ matters greatly. If $Z_\mathrm{gr}=10^{-8}$ a large portion of the atmosphere is isothermal, causing an exponential rise of the gas density once inside the Bondi radius ($r_\mathrm{Bondi}\equiv GM_\mathrm{core}/(k_B T_\mathrm{disk}/\mu)$ indicates the point where the escape velocity of the planet equals the thermal velocity of the gas). At a certain point the gas opacity will become larger than $\kappa_\mathrm{gr}$. This is indicated by a circle. Further in, at a gas density $\approx$10$^6\ \rho_\mathrm{disk}$, the atmosphere becomes convective (the RCB: triangle) and $T$ and $\rho_\mathrm{gas}$ become power-laws. For the $Z_\mathrm{gr}=10^{-2}$ run this transition occurs much higher in the atmosphere. There is no large isothermal outer layer and, consequently, the atmosphere mass is much smaller (\Tb{runs}).

\Fg{cool}b presents the grain abundance, size, and opacity. Because $s$ and $Z_\mathrm{gr}$ are fixed there is little structure. Note the increase in $\kappa_\mathrm{gr}$ for smaller $r$. As the radiation peak shifts to shorter wavelength due to higher temperature the grains become optically larger.

\subsection{Including grain growth}
Next, we include \eqs{master-4a}{master-4b} with constant $\Mdotcum{dep}(r)$. The grain aerodynamical properties are reflected in their stopping time: $t_\mathrm{stop}=mv_\mathrm{gas}/F_\mathrm{drag}$, with $v_\mathrm{gas}$ the particle-gas velocity and $F_\mathrm{drag}$ the drag force \citep[][]{Weidenschilling1977}. Generally, $t_\mathrm{stop}$ is found iteratively as it depends on (the settling) velocity and the settling velocity on $t_\mathrm{stop}$: $v_\mathrm{settl}=g_r t_\mathrm{stop}$, where $g_r=GM_\mathrm{core}/r^2$ is the local gravitational acceleration.

We include two relative velocity sources. The first is Brownian (thermal) motions, $\Delta v_\mathrm{bm}=\sqrt{16 k_B T_\mathrm{gas}/\pi \mast}$ (for equal-size particles). The second is differential drift motions $\Delta v_\mathrm{dd}$ that arise due to settling. As settling velocities are the same for identical particles, growth depends on the width of the size distribution. We parametrize this effect by a parameter $\chi$ ($<$1) such that $\Delta v_\mathrm{dd} \equiv \chi v_\mathrm{settl}$. We use $\chi=0.1$ \citep{OkuzumiEtal2011}.

The growth rates arising from Brownian motion and differential drift (settling) are given by $T_{\mathrm{growth},i}^{-1}=(n_\mathrm{gr}\sigma_\mathrm{gr}\Delta v_i)=3Z_\mathrm{gr}\rho_\mathrm{gas} \Delta v_i/\rho_\bullet s$ where $\Delta v_i$ is either $\Delta v_\mathrm{bm}$ or $\Delta v_\mathrm{dd}$. We simply add these two rates. Under most conditions, differential drift dominates.

We fix $Z_\mathrm{gr}=10^{-2}$ at the outer boundary, which implies a mass flux of $\Mdotcum{dep} =\Mdotcum{disk} = 5\times10^{-9}\ M_\oplus\ \mathrm{yr}^{-1}$ in \micr-size grains. This is 0.05\% of the total mass flux (\fg{deposit}). 

\begin{figure*}
  \centering
  \includegraphics[width=180mm]{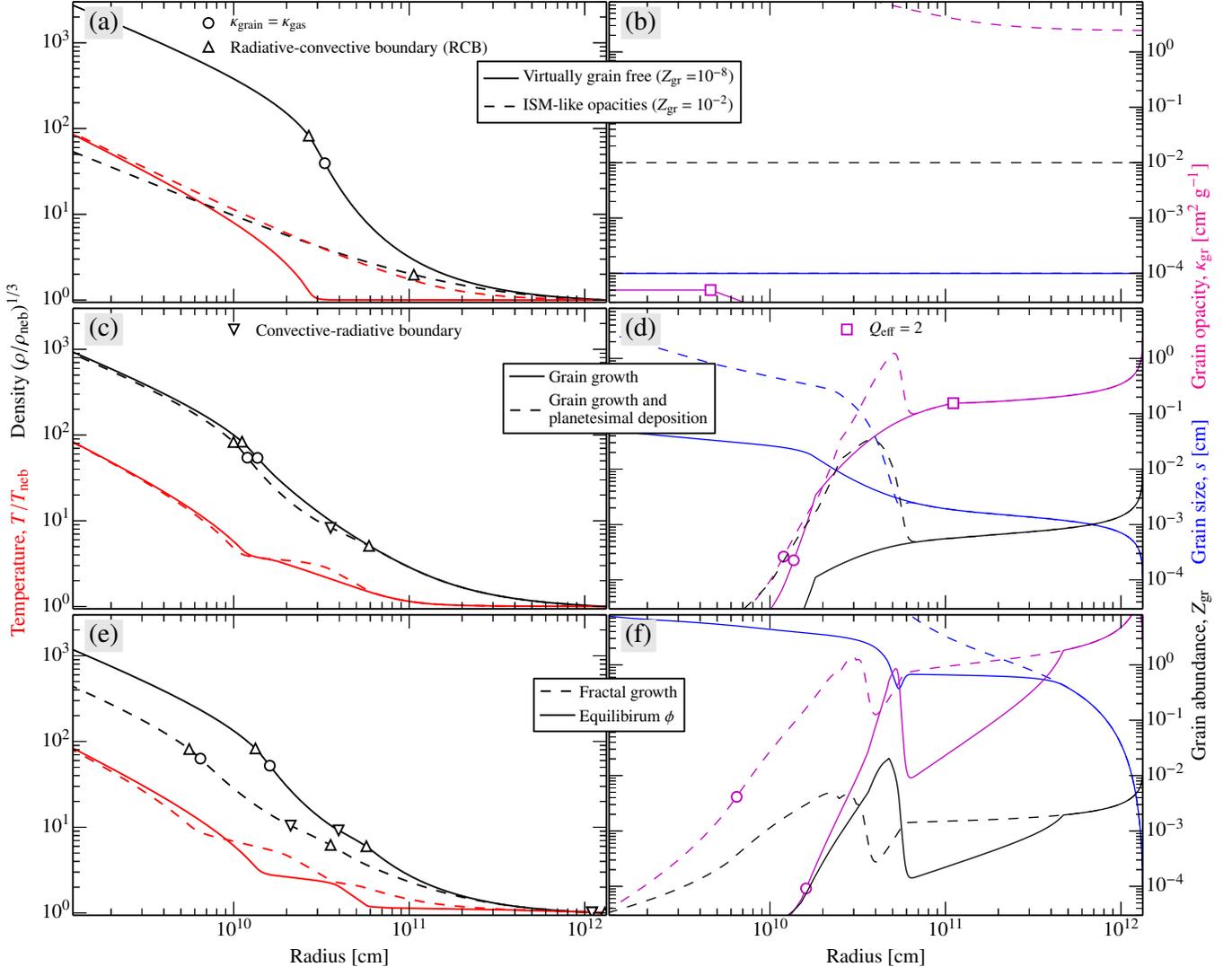}
  \caption{Atmosphere structure profiles for several grain growth scenarios. Left panels show the temperature and density as function of radius.  Right panels show the \textit{corresponding} grain properties: the grain size, the grain abundance, and the grain opacity. Curves of the same linestyle (solid, dashed) belong to the same run. \textit{Top panels}: no grain growth at constant grain abundance of $Z_\mathrm{gr}=10^{-8}$ (solid) and $Z_\mathrm{gr}=10^{-2}$ (dashed). \textit{Middle panels}: including grain growth (solid) and grain growth and deposition (dashed). \textit{Lower panels}: using a fractal law for the grain filling factor (dashed) and using the equilibrium filling factor (solid).}
  \label{fig:cool}
\end{figure*}

Figures \ref{fig:cool}c and d (solid lines) present the results. The micron-size grains that enter the atmosphere from the disk quickly coagulate to sizes $\sim$10 \micr, providing an immediate drop in the grain abundance as the settling velocity increases. Obviously, a more self-consistent model would already account for the grain evolution that takes place in the parent disk\citep[\eg][]{ZsomEtal2011,BirnstielEtal2012i}.  In the outer layers of the atmosphere competing mechanisms keep $Z_\mathrm{gr}$ and $\kappa_\mathrm{gr}$ relatively constant: (with decreasing $r$) $g_r$ increases but $t_\mathrm{stop}$ decreases due to the higher densities. Also, the grain  efficiency $Q_\mathrm{eff}$ increases, until the point where $Q_\mathrm{eff}=2$ is reached (square). The `knee' seen at $r=2\times10^{10}$ cm results from the transition from Epstein to Stokes drag, which boosts the settling velocity.

%As a result, the grain opacity is strongly reduced with respect to the `ISM-case'. 
The decrease of $\kappa_\mathrm{gr}$ with decreasing $r$ is essential for prolonging the extent of the radiative zone, where density gradients are much steeper. The RCB is determined by the gas opacity -- a result valid for all models that employ grain growth. The shape of the temperature and density structures bear closer resemblance to the virtually grain-free ($Z_\mathrm{gr}=10^{-8}$) case than to the ISM opacity ($Z_\mathrm{gr}=10^{-2}$). The atmosphere masses lie in between these limiting models (\Tb{runs}).

\subsection{Including mass deposition}
\label{sec:depos}
The above models assumed that the planetesimals accreted by the core remained intact until they hit the core, where they liberated most of their binding energy. We next consider planetesimals that disintegrate in the atmosphere.

This means that a deposition profile must be specified. We choose:
\begin{equation}
  -\dMdrF{dep}
  = \rho_\mathrm{gas} \frac{d\dot{M}_\mathrm{dep}}{d\Sigma}
  = \rho_\mathrm{gas} \Mdottot{plts} P_\mathrm{ln}(\Sigma;\Sigma_\mathrm{crit},\sigma)
  \label{eq:dMdrF}
\end{equation}
where $\Sigma$ is the column density as measured from the top of the atmosphere, $\Mdottot{plts}=\Mdottot{tot} -\Mdottot{disk}$, and $P_\mathrm{ln}$ is the log-normal distribution:
\begin{equation}
  P_\mathrm{ln}(x;\mu,\sigma)
  = \frac{1}{\sigma x \sqrt{2\pi}} \exp \left[ -\frac{1}{2\sigma^2} \left( \log (x/\mu) \right)^2 \right],
  \label{eq:Psig}
\end{equation}
with $\sigma$ controlling the width of $P_\mathrm{ln}(\Sigma)$, a proxy for the dispersion in the planetesimal sizes.  This distribution is chosen purely for mathematical convenience. 
%The critical column density $\Sigma_\mathrm{crit}$ may be taken as the point where the planetesimals have encountered their own mass $\approx$$\rho_\mathrm{\bullet}s_\mathrm{plts}$. 
For illustrative purposes, we choose a very low value for $\Sigma_\mathrm{crit}$ to ensure that grains are deposited in the radiative zone. 
%Our adopted values for $\Sigma_\mathrm{crit}$ and $\sigma$ are arbitrary and
%The shape parameter $\sigma$ controls the width of the distribution, which is a proxy for the size distribution of the planetesimals. 
\Fg{deposit} illustrates the differential (\eq{dMdrF}) and cumulative deposition profiles. The latter also includes the disk contribution, $\Mdotcum{disk}$.

%Finally, we have calculated the spatial density of the $m_\mathrm{dep}$-particles (\se{bimodal}) and included the resulting opacity in the \xxx.
There is, then, a 2,000-fold increase in $\dot{M}_\mathrm{dep}$ around a column density $\Sigma_\mathrm{crit}$. Does all of this matter? Scarcely. \Fg{cool}c shows that the profiles including deposition (dashed) hardly deviate from the profiles without deposition (solid). The combined process of grain coagulation and grain settling provide a powerful antidote against the increased grain abundance. This is illustrated in \fg{cool}d. At the point where the injection takes place ($r\approx5\times10^{10}$ cm) grains start to grow rapidly. This has two key effects: (i) a decreasing opacity per unit grain mass; (ii) a lower grain abundance due to the increased settling velocity. Together, they act to suppress the grain opacity (magenta line): the increase in $\kappa_\mathrm{gr}$ is limited to a narrow -- convective -- shell but does not propagate deeper into the atmosphere.

When accounting for the bimodal correction (\se{bimodal}) we find that the `opacity bump' increases by $\approx$5, but that it does not affect the above conclusions.

\subsection{Grain internal composition}
In the above runs we assumed that the internal density of the grains equals that of the monomers: $\rho_\bullet=\rho_\mathrm{0}=3\ \mathrm{g\ cm^{-3}}$. However, the initial stages of grain growth are characterized by the emergence of agglomerates \citep{OrmelEtal2007,OkuzumiEtal2009} where the filling factor ($\phi=\rho_\bullet/\rho_0$) decreases with size. Let us assume a fractal law for the filling factor: $\phi=\phi_\mathrm{frac}=(s_0/s)^\delta$ where $\delta=0$ corresponds to compact coagulation ($\phi=1$) and $\delta=1$ to 2D structures (pancakes) where surface area ($s^2$) is proportional to mass ($\sim$$s^3\phi$).  See \citet{OkuzumiEtal2009,OkuzumiEtal2012} for physical models for the evolution of $\phi$.

The dashed lines in \fg{cool}e,f show the result for $\delta=0.8$. As particles become very fluffy, their settling is suppressed and their abundance increases. The growth is dramatic: the structures easily reach sizes of meter-to-kilometers (this result is extremely sensitive to the adopted value of $\delta$). Deposition of grains by planetesimals is in the case of fractal growth more permanent: note the broadening in $\kappa_\mathrm{gr}$ at small $r$. As a result of this pileup the grain opacity is larger, which suppresses the gas density compared to the compact growth models.

However, the existence of such fluffy particles is questionable, as they compact collisionally \citep{DominikTielens1997,WadaEtal2008} and by gas drag. Recently, \citet{KataokaEtal2013} argued that the compressive strength of a highly-porous particle is on the order of $\phi^3 E_\mathrm{roll}/s_0^3$, where $E_\mathrm{roll}$ is the energy needed to move two grains in contact over an angle of 90 degrees. Equating this internal strength to the pressure experienced by gas drag, $P_\mathrm{gas}= mv/\pi s^2 t_\mathrm{stop}$ one retrieves the equilibrium filling factor \citep{KataokaEtal2013i}:
\begin{equation}
  \phi_\mathrm{eq} 
  = \left( \frac{4\rho_0s_0^3 s v_\mathrm{settl}}{3E_\mathrm{roll} t_\mathrm{stop}} \right)^{1/2}
  = \left( \frac{4\rho_0 s_0^3 s g_r}{3E_\mathrm{roll}} \right)^{1/2}.
\end{equation}
This expression shows that $\phi_\mathrm{eq}$ always increases with decreasing $r$ as long as $s$ increases: the agglomerates compact.

Figures \ref{fig:cool}e and f (solid curves) express this point. We have adopted $\phi=\max(\phi_\mathrm{eq},\phi_\mathrm{frac})$ and a laboratory-measured value for the rolling energy \citep{HeimEtal1999}. Initially, because $\phi_\mathrm{eq}<\phi_\mathrm{frac}$ the growth is fractal and the solid and dashed curves coincide. Very quickly, however, static compression due to gas drag compacts the grains. The grain size then stabilizes at $\sim$mm, until the influx of fresh grains due to planetesimal deposition causes a sharp rise in $s$. On average, the combined effects of the initial fractal growth followed by compaction suppresses the grain opacities with respect to the compact growth cases (\fg{cool}c,d), resulting in large atmosphere masses (\Tb{runs}).

\section{Summary}
\label{sec:summary}
Perhaps the most striking feature of our calculations is that the grain opacity $\kappa_\mathrm{gr}$ is so little affected by the influx of material. The 2,000-fold increase in $\Mdotcum{dep}$ due to planetesimal breakup and fragmentation, which was modeled as 100\% efficient in its conversion to micron-size grains, left few imprints to the atmospheric structures. The added surface material simply coagulates away.

This conclusion agrees with \citet{MovshovitzPodolak2008}.  In a contemporaneous manuscript, \citet{Mordasini2014} investigated this aspect in detail. By equating the grain settling timescale to $T_\mathrm{grow}$, he obtained analytical expressions for $\kappa_\mathrm{gr}$ and found that it is independent of the grain abundance or the solid mass flux $\dot{M}_\mathrm{dep}$. Both \citet{Mordasini2014} and this work arrive at the conclusion that grain-free opacities are more relevant than ISM-like opacities. 

Nevertheless, it is useful to follow the grain opacity with methods presented in this work for two reasons. First, it is hard to predict \textit{a priori} the appropriate value -- let alone the profile -- for $\kappa_\mathrm{gr}$ (or $Z_\mathrm{gr}$) in a convoluted and time-dependent environment. Second, our method is computationally cheap: just one additional atmospheric structure equation.%to integrate.

\acknowledgments
I acknowledge support for this work by NASA through Hubble Fellowship grant \#HST-HF-51294.01-A awarded by the Space Telescope Science Institute, which is operated by the Association of Universities for Research in Astronomy, Inc., for NASA, under contract NAS 5-26555. I appreciate exchanges with the referee, T. Birnstiel, A. Youdin and the willingness of C. Mordasini to share an early draft of his manuscript.

\bibliographystyle{apj}
%\bibliography{ads,mybibl}
%\end{document}

\end{document}